# Radiation-hard ASICs for optical data transmission in the ATLAS pixel detector

K.K. Gan, K.E. Arms, M. Johnson, H. Kagan, R. Kass, C. Rush, S. Smith,
R. Ter-Antonian, M.M. Zoeller

Department of Physics, The Ohio State University, Columbus, OH 43210, USA

A. Ciliox, M. Holder, M. Ziolkowski

Fachbereich Physik, Universitaet Siegen, 57068 Siegen, Germany

We have developed two radiation-hard ASICs for optical data transmission in the ATLAS pixel detector at the LHC at CERN: a driver chip for a Vertical Cavity Surface Emitting Laser (VCSEL) diode for 80 Mbit/s data transmission from the detector, and a Bi-Phase Mark decoder chip to recover the control data and 40 MHz clock received optically by a PIN diode. We have successfully implemented both ASICs in 0.25 µm CMOS technology using enclosed layout transistors and guard rings for increased radiation hardness. We present results from circuits of final design and from irradiation studies with 24 GeV protons up to 62 Mrad ($2.3 \times 10^{15}$ p/cm$^2$).

## 1. INTRODUCTION

The ATLAS pixel detector [1] consists of two barrel layers and two forward and backward disks which provide at least two space point measurements. The pixel sensors are read out by front-end electronics controlled by the Module Control Chip (MCC). The low voltage differential signal (LVDS) from the MCC is converted by the VCSEL Driver Chip (VDC) into a single-ended signal appropriate to drive a VCSEL. The optical signal from the VCSEL is transmitted to the Readout Device (ROD) via a fiber.

The 40 MHz beam crossing clock from the ROD, bi-phase mark (BPM) encoded with the data (command) signal to control the pixel detector, is transmitted via a fiber to a PIN diode. This BPM encoded signal is decoded using a Digital Opto-Receiver Integrated Circuit (DORIC). The clock and data signals recovered by the DORIC are in LVDS form for interfacing with the MCC.

The ATLAS pixel optical link contains 448 VDCs and 360 DORICs with each chip having four channels. The chips will be mounted on 180 chip carrier boards (opto-boards). The optical link circuitry will be exposed to a maximum total fluence of $2 \times 10^{15}$ 1-MeV $n_{eq}$/cm$^2$ during ten years of operation at the LHC. In this paper we describe the development of the radiation-hard VDC and DORIC.

## 2. VDC CIRCUIT

The VDC converts an LVDS input signal into a single-ended signal appropriate to drive a VCSEL in a common cathode array. The output current of the VDC is to be variable between 0 and 20 mA through an external control current, with a standing current (dim current) of ~1 mA to improve the switching speed of the VCSEL. The rise and fall times of the VCSEL driver current are required to be less than 1 ns. In order to minimize the power supply noise on the opto-board, the VDC should also have constant current consumption independent of whether the VCSEL is in the bright (on) or dim (off) state.

Figure 1 shows a block diagram of the VDC circuit. An LVDS receiver converts the differential input into a single-ended signal. The differential driver controls the current flow from the positive power supply into the anode of the VCSEL. The VDC circuit is therefore compatible with a common cathode VCSEL array. An externally controlled voltage, $V_{Iset}$, determines the current $I_{set}$ that sets the amplitude of the VCSEL current (bright minus dim current), while an externally controlled voltage, tunepad, determines the dim current. The differential driver contains a dummy driver circuit which in the VCSEL dim state draws an identical amount of current from the positive power supply as is flowing through the VCSEL in the bright state. This enables the VDC to have constant current consumption.

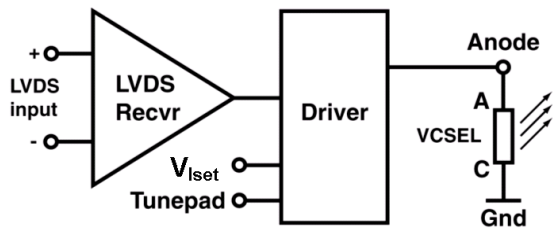

Figure 1: Block diagram of the VDC circuit.

## 3. DORIC CIRCUIT

The DORIC decodes BPM encoded clock and data signals received by a PIN diode. The BPM signal is derived from the 40 MHz beam crossing clock by sending only transitions corresponding to clock leading edges. In the absence of data bits (logic level 1), this results simply in a 20 MHz signal. Any data bit in the data stream is encoded as an extra transition at the clock trailing edge.

The amplitude of the current from the PIN diode is expected to be in the range of 40 to 1000 µA. The 40 MHz clock recovered by the DORIC is required to have a duty cycle of (50 ± 4)% with a total timing error of less than 1 ns. The bit error rate of the DORIC is required to be less than $10^{-11}$ at end of life.

Figure 2 shows a block diagram of the DORIC circuit. In order to keep the PIN bias voltage (up to 10 V) off the DORIC, we employ a single-ended preamp circuit to amplify the current produced by the PIN diode. Since single-ended preamp circuits are sensitive to power supply noise, we utilize two identical preamp channels: a signal channel and a noise cancellation channel. The signal channel receives and amplifies the input signal from the anode of the PIN diode, plus any noise picked up by the circuit. The noise cancellation channel amplifies noise similar to that picked up by the signal channel. This noise is then subtracted from the signal channel in the differential gain stage. To optimise the noise subtraction, the input load of the noise cancellation channel should be matched to the input load of the signal channel (PIN capacitance) via an external dummy capacitance.

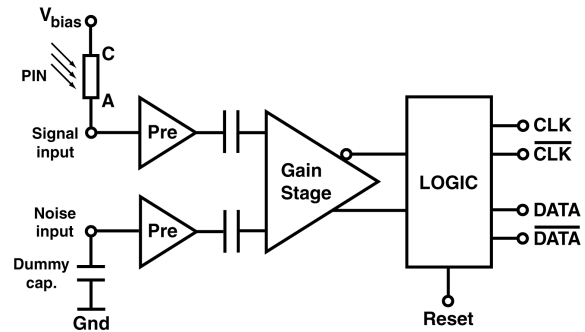

Figure 2: Block diagram of the DORIC circuit.

## 4. RESULTS FROM IBM 0.25 MICRON SUBMISSIONS

The pixel detector design of the VDC and DORIC takes advantage of the development work for similar circuits [2] used by the outer detector, the SemiConductor Tracker (SCT). Both SCT chips attain radiation-tolerance by using bipolar integrated circuits (AMS 0.8 µm BICMOS) and running with high currents in the transistors at 4 V nominal supply voltage. These chips are therefore not applicable for the higher radiation dosage and lower power budget requirements of the pixel detector.

We originally implemented the VDC and DORIC in radiation-hard DMILL 0.8 µm technology with a nominal supply voltage of 3.2 V. An irradiation study of the DMILL circuits in April 2001 with 24 GeV protons at CERN showed severe degradation of circuit performance. We therefore migrated the VDC and DORIC designs to the standard deep submicron (0.25 µm) CMOS technology which had a nominal supply voltage of 2.5 V. Employing enclosed layout transistors and guard rings [3], this technology was expected to be very radiation hard. We had five prototype runs using 3-metal layouts over the course of two years with IBM as the foundry. For the engineering run, the layouts were converted to 5-metal layouts in order to share the wafers with the MCCs for cost saving.

We have extensively tested the chips to verify that they satisfy the design specifications. Figure 3 shows the VCSEL current generated by the VDC as a function of the external control current $I_{set}$. The saturation at high $I_{set}$ is due to the large serial resistance of the VCSEL. The dim current is close to the design value of 1 mA. The performance of the chips on opto-boards has been studied in detail. Each opto-board contains seven optical links. The typical PIN current thresholds for no bit errors are low, ~15-40 μA.

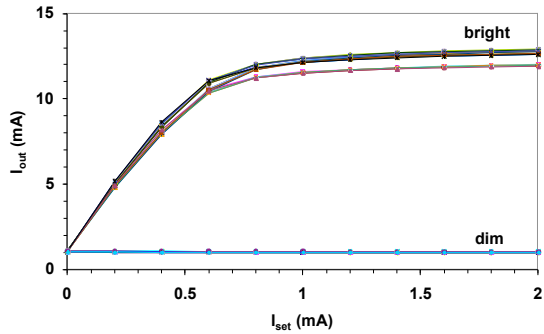

Figure 3: VCSEL drive current vs. $I_{set}$ of eight 4-channel VDCs.

## 5. IRRADIATION STUDIES

In the last three years, we have performed three irradiations of the VDCs and DORICs produced using the deep submicron process. We used 24 GeV protons at CERN for the study. No significant degradation in the performance of the chips was observed. In the following we describe the results from the last irradiation in more detail.

We use two setups during the irradiation. In the first setup ("cold box"), we perform electrical testing of VDCs and DORICs. No optical components (VCSEL or PIN) are used, allowing direct study of the possible degradation of the chips without additional complications from the optical components. For the VDCs, we monitor the rise and fall times and the bright and dim currents. For the DORICs, we monitor the minimum input signal for no bit errors, clock jitter and duty cycle, rise and fall times and amplitude and average of the clock and command LVDS.

Four DORICs and four VDCs were irradiated in the cold box setup. We observe no significant degradation in the chip performance up to a total dose of ~62 Mrad, except the average of the clock LVDS of one DORIC, which increases by 10%, an acceptable change. Some examples of the comparisons of the performances before and after irradiation are shown below. Figure 4 shows a comparison the bright and dim currents of the VDCs. Figure 5 shows a comparison of the rise and fall times of the decoded clock of the DORICs.

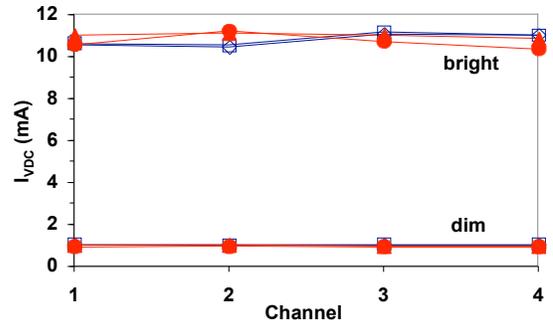

Figure 4: VCSEL drive currents of two 4-channel VDCs before (open) and after (filled) irradiation.

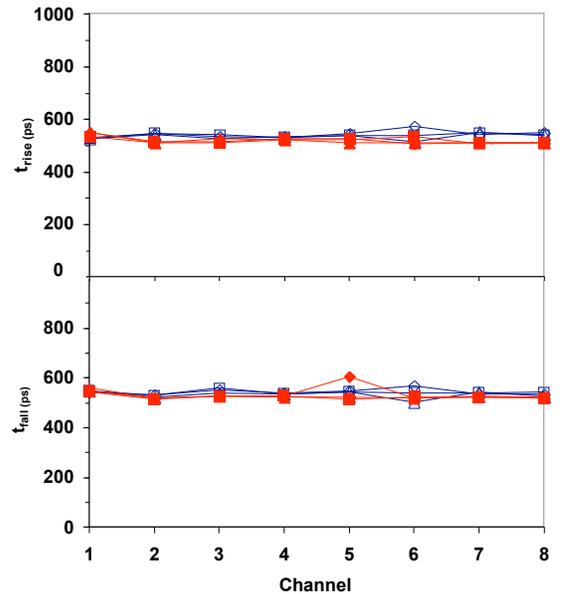

Figure 5: Rise and fall times of the decoded clock of two 4-channel DORICs before (open) and after (filled) irradiation.

In the second setup ("shuttle"), we test the performance of the opto-link using opto-boards. In the control room, we generate bi-phase mark encoded pseudo-random signals for transmission via 25 m of optical fibers to the opto-boards. The PIN diodes on the opto-boards convert the optical signals into electrical signals. The DORICs then decode the electrical signals to extract the clock and command LVDS. The LVDS are fed into the VDCs and converted into signals that are appropriate to drive the VCSELs. The optical signals are then sent back to the control room for comparison with the generated signals. We remotely move the opto-boards on the shuttle out of the beam to anneal the VCSELs. We typically irradiate the opto-boards for 5 hours (~5 Mrad) and then anneal the VCSELs for the rest of the day with large current (~13 mA in the third irradiation).

We irradiated four opto-boards with the dosage of (8-10) x $10^{14}$ p/cm$^2$ (22-28 Mrad). We observe that the PIN current thresholds for no bit errors are all below 55 µA and remain constant or decrease slightly up to the total dosage of 25 Mrad as shown in Fig. 6.

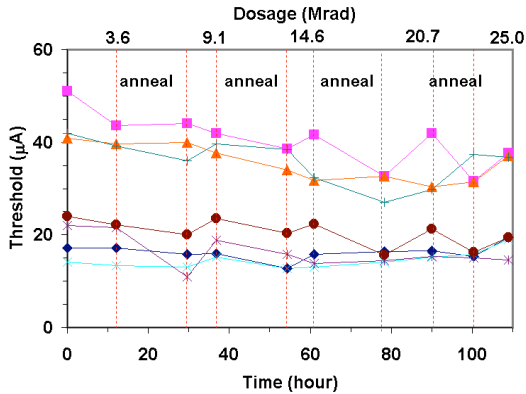

Figure 6: PIN current thresholds as a function of time (dosage) for one of the opto-boards with seven active links in the shuttle setup.

The optical power from the opto-boards were monitored in the control room during the irradiation. Figure 7 shows the optical power as a function of time (dosage) for one of the opto-boards. We observe a general trend in the data: during the irradiation the optical power decreases; the optical power increases during the annealing, as expected. The power loss is due to radiation damage to the VCSELs as the VDCs and DORICs show no radiation damage with up to ~62 Mrad of irradiation. The power loss is significant due to insufficient time for adequate annealing. We received the opto-boards from CERN a month after irradiation and have measured their properties. All VCSEL channels were working but one channel stopped producing light after a few hours of annealing. After ~120 hours of annealing at the VCSEL current of ~13 mA, the optical power of all working links were above 350 µW, the specification for absolute minimum power after irradiation.

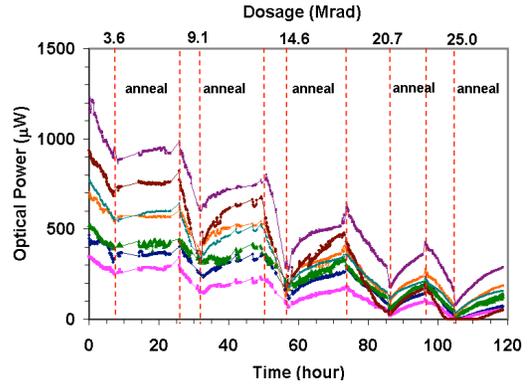

Figure 7: Optical power as a function of time (dosage) in the data channels for one of the opto-boards with seven active links in the shuttle setup.

## 6. SUMMARY

We have developed VDCs and DORICs in deep submicron (0.25 µm) technology using enclosed layout transistors and guard rings for improved radiation hardness. The circuits meet all the requirements for operation in the ATLAS pixel optical link and further appear to be sufficiently radiation hard for ten years of operation at the LHC.

## ACKNOWLEDGEMENTS

This work was supported in part by the U.S. Department of Energy under contract No. DE-FG-02-91ER-40690 and the German Federal Minister for Research and Technology (BMBF) under contract 056Si74.